\documentclass{llncs}
\usepackage[hyphens]{url}
\usepackage[hidelinks]{hyperref}

\usepackage[utf8]{inputenc}
\usepackage{subcaption}

\usepackage{mathabx}
\usepackage{microtype}

\usepackage{multirow}

\newcommand\Small{\fontsize{9}{9.2}\selectfont}
\newcommand*\LSTfont{%
	\Small\ttfamily\SetTracking{encoding=*}{-60}\lsstyle}

\usepackage{listings}
\usepackage{tikz}
\newcommand*\circled[1]{\tikz[baseline=(char.base)]{\node[shape=circle,draw,inner sep=1pt] (char) {#1};}}

\DeclareRobustCommand\circledr[1]{\tikz[baseline=(char.base)]{\node[shape=circle,draw,inner sep=1pt] (char) {#1};}}

\begin{document}

\title{An Easy \& Collaborative RDF Data Entry Method using the Spreadsheet Metaphor}
\author{Markus Schröder\inst{1} \and Christian Jilek\inst{1} \and Jörn Hees\inst{1} \and\\Sven Hertling\inst{2} \and Andreas Dengel\inst{1,3}}

\institute{German Research Center for Artificial Intelligence (DFKI) GmbH\\Trippstadter Stra{\ss}e 122, 67663 Kaiserslautern, Germany,\\
\email{\{markus.schroeder, christian.jilek, joern.hees, andreas.dengel\}@dfki.de},
\and
Data and Web Science Group, University of Mannheim, Germany,\\
\email{sven@informatik.uni-mannheim.de},
\and
Knowledge-Based Systems Group, Department of Computer Science,\\TU Kaiserslautern, P.O. Box 3049, 67653 Kaiserslautern, Germany\\
}
\maketitle

\begin{abstract}

Spreadsheets are widely used by knowledge workers, especially in the industrial sector.
Their methodology enables a well understood, easy and fast possibility to enter data.
As filling out a spreadsheet is more accessible to common knowledge workers than defining RDF statements, in this paper, we propose an easy-to-use, zero-configuration, web-based spreadsheet editor that simultaneously transfers spreadsheet entries into RDF statements.
It enables various kinds of users to easily create semantic data whether they are RDF experts or novices.
The typical scenario we address focuses on creating instance data starting with an empty knowledge base that is filled incrementally.
In a user study, participants were able to create more statements in shorter time, having similar or even significantly outperforming quality, compared to other approaches.

\end{abstract}

\section{Motivation}
One of the first tasks in our industry projects is to gather knowledge about the customer's domain, preferably in a semantic representation like RDF.
In order to extract domain knowledge, data dumps provided by the customer are usually analysed, which is often not enough: interrelations between different parts of the data are often missing, data may be incomplete or misleading, information is not available in written form, but only in people's minds, etc.
For us, especially such tacit knowledge is often valuable, requiring us to also focus intensively on the domain experts.
Instead of rather passively involving our partners, e.g. by doing interviews (often costly), we prefer to enable them to actively and directly communicate their expertise using given software.
Available tools (e.g., the frequently used Protégé\footnote{\url{http://protege.stanford.edu}}) require a certain training period, as well as considerable RDF knowledge, in order to be used (and configured).
In contrast, the spreadsheet metaphor is widely known and well suited for enabling \textit{all kinds} of users to manually enter data.
Thus, our goal is to find a balance between the two.
We therefore propose an easy-to-use, zero-configuration\footnote{using a fixed mapping, see Section \ref{sec:approach}} spreadsheet editor that simultaneously transfers spreadsheet entries into RDF statements.
The typical scenario we address focuses on creating instance data (ABox) starting with an empty knowledge base that is filled incrementally.
Collecting all relevant information is usually an iterative process, requiring the collaboration of several experts, as well as the occasional assistance of knowledge engineers.
The task of the latter is to possibly provide initial guidance, intermediate feedback about modelling consequences, or doing data cleanup.
Since such a modelling process involves a lot of communication (expert-to-expert, expert-to-knowledge
engineer, knowledge engineer-to-knowledge engineer), all contributors should immediately see updates by others.
We therefore designed our spreadsheet editor to be web-based, i.e. hosted on the intranet – or the internet if confidentiality requirements permit this.
The application, which was briefly demonstrated\footnote{\url{http://www.dfki.uni-kl.de/~mschroeder/rdf-spreadsheet-editor/}} in \cite{schroder2017rdf}, is evaluated in a user study which showed that participants were able to easily create meaningful triples although being rather inexperienced in RDF.\\

This paper is structured as follows:
Our application is presented in Section \ref{sec:approach}, followed by the results of the aforementioned user study (Section \ref{sec:eval}).
In Section \ref{sec:relwork}, a detailed overview of related work in this area is presented.
Last, we conclude this paper by giving a short summary and outlook on possible future work in Section \ref{sec:concl}.

\section{Approach}
\label{sec:approach}
Our approach is a web-based spreadsheet editor, that simultaneously transfers spreadsheet inputs to RDF statements and also adds them to a knowledge base.
By this we enable users to enter (and modify) semantic data in a familiar way.
Although there exist several importing tools, we decided against them to better suite the interactive and communication-intensive nature of our targeted scenarios.
We use a simple and fixed \textit{class per sheet} and \textit{entity per row} mapping similar to csv2rdf\footnote{\url{https://www.w3.org/TR/csv2rdf/}}.
This is a trade-off decision favouring convenient manual data entry over highly configurable data import, since we primarily focus on gathering the knowledge in people's minds.
The main features are creating and manipulating RDF classes, properties, instances and assertions.
As stated in the last section, our goal was not to provide a full-featured ontology editor, but to focus on easily creating instance data (ABox).
Especially, our tool supports the user by automatically inferring and creating domain and range statements or auto completion of resource labels, for example.
Common vocabularies (like FOAF) can simply be added in order to reuse their classes and properties (i.e., referencing them in spreadsheets).
Knowledge bases can be imported and exported using typical triple data formats.
For a first impression we kindly refer the reader to our application's demo\footnote{\url{http://www.dfki.uni-kl.de/~mschroeder/rdf-spreadsheet-editor/}} \cite{schroder2017rdf}.\\

Using an example, we will present the basic functionality (Section \ref{sec:basic}), as well as extensions that further ease the usage of our tool (Section \ref{sec:ext}).

\subsection{Basic Functionality}
\label{sec:basic}

\begin{figure}[h]
	\centering
	\includegraphics[width=0.8\textwidth]{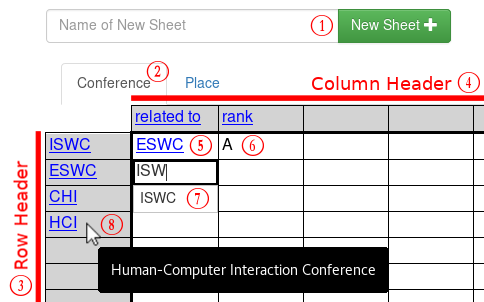}
	\caption{GUI of our RDF spreadsheet editor with already entered conferences and additional information. Feature remarks: \circledr{1} \circledr{2} create a new sheet, \circledr{3} instances, \circledr{4} properties, \circledr{5} object statement, \circledr{6} literal statement, \circledr{7} auto complete feature and \circledr{8} resource's comment when mouse hovering.}
	\label{demo}
\end{figure}

To show the basic functionality of our app, we use the example of collecting information about conferences in a spreadsheet.
Its graphical user interface is depicted in Figure \ref{demo} (we annotated the main features in red).
Like it is common for spreadsheet editors, users may create several sheets.
Naming a sheet \circled{1} creates a new class \circled{2}.
The row header \circled{3} is used to create resources which are instances of this class.
Entering a label in a column header \circled{4} generates a new property having the class as its domain.
Cell \circled{5} instantiates a resource labelled $ESWC$ and links it to the resource $ISWC$ (row header) using the $related~to$ property (column header) resulting in the following triple statement: ($ISWC$, $related~to$, $ESWC$).
Forcing the application to create a literal \circled{6} instead of a resource is done by prepending a single quotation mark as common in spreadsheet tools indicating text\footnote{Note that the prepending a single quotation mark is not shown in Figure \ref{demo} because the cell is not currently being edited.}.
There is an auto completion feature \circled{7} which operates on the resources' text literals.
Using a text area (not visible in the screen shot) a comment for each resource can be stated, which is then presented when mouse hovering it \circled{8}.

In the following, we will have a closer look at the generated RDF triple statements.
For each new resource a URI is generated having a random UUID as postfix.
The entered text in the cell is used as the resource's label and also serves as an ``identifier'' as explained in more detail in the next section.
Since English is the selected language in our example, the label's language tag is set accordingly.
We apply a \textit{class per sheet} mapping, thus entering a name in the row header \circled{3}, e.g. \textit{ISWC}, creates a resource which is an instance of \textit{Conference}.
Column header entries \circled{4} like \textit{related~to} lead to the creation of properties having this sheet's class \textit{Conference} as its \texttt{rdfs:domain}.
Entering \textit{ESWC} into the cell that intersects the \textit{ISWC} row and the \textit{related~to} column \circled{5} triggers two actions:
First, the \textit{ESWC} resource is created, then a statement \textit{ISWC related to ESWC} is generated.
Here we apply an \textit{entity per row} mapping.
Note that this is not dependent on the order in which the cells are populated.
A user may first fill the (intersecting) cell, then the row and column headers in order to create the same statement.
To distinguish between resources and literals, one has to prepend a single quote in order to make the cell content a literal.
In this example the user would like to save the \textit{rank}\footnote{The rank value is taken from \url{http://portal.core.edu.au/conf-ranks/1338/}} of \textit{ISWC} as a language string.
Thus, \texttt{'A} is entered in \circled{6} (the single quote is not visible in the screen shot, since the cell is currently not in edit mode).
Since the \textit{rank} column currently only contains \texttt{rdf:langString} values (i.e., \texttt{"A"@en}), its \texttt{rdfs:range} is inferred accordingly.
The resulting RDF statements are shown in Listing \ref{listingforall}.

\begin{lstlisting}[caption=The resulting triples for our example (all URIs are shortened for the sake of readability and have an exemplary \textit{ex} namespace), label=listingforall]
ex:1cfd a           rdfs:Class ;
        rdfs:label  "Conference"@en .      # (2)
                
ex:99f2 a           ex::1cfd , owl:Thing ;
        rdfs:label  "ISWC"@en ;            # (3)
        ex:ccf1     ex:76b9 ;              # (5)
        ex:6942     "A"@en .               # (6)
                
ex:ccf1 a           rdf:Property ;
        rdfs:domain ex:1cfd ;
        rdfs:label  "related to"@en .      # (4)
                
ex:76b9 a           owl:Thing ;
        rdfs:label  "ESWC"@en .
                
ex:6942 a           rdf:Property ;
        rdfs:label  "rank"@en ;
        rdfs:domain ex:1cfd ;
        rdfs:range  rdf:langString .
\end{lstlisting}

\subsection{Extensions}
\label{sec:ext}
Using the previously presented basic functionality it is already possible to make simple RDF statements.
However, to simplify the usage of our tool we implemented several extensions.

Making it possible to address resources by their label, we have to disambiguate them.
Thus, entering \textit{ISWC}, for example, does not create a new resource, but reuses the existing one.
This feature may be disabled if desired.
Note that in our targeted, rather small closed-world scenarios ambiguities are very rare and can easily be resolved (e.g., assigning a slightly different label).
To explicitly refer to an existing resource, users then have two possibilities:
using the previously introduced (1) auto completion feature, which shows a list of suggestions while typing.
If one of them is selected, the resource's URI instead of its label is used.
Additionally, it is possible to (2) copy \& paste a resource from one cell to another.
Again, the system uses the resource's URI instead of its label.
Reusing the resources throughout one or on different sheets enables users to link instances (ABox) and create a more interconnected RDF graph.
Additionally, other users can benefit from classes and properties (TBox) that have already been modelled by their colleagues.
Let us consider a short example.
While it is enough to say that a \textit{Conference} \textit{CHI} \textit{is located in} a place \textit{Denver} in one sheet, details about this location may be given in another.
This can easily be done by just adding \textit{Denver} to this new sheet's row header and providing more properties using its columns.

Apart from a label, each resource can be equipped with a comment, which is particularly helpful if two resources share the same name:
For example, \textit{HCI} could be the name of a research area or a conference.
If a resource is in focus, it may be commented on using a text area at the bottom of the screen.
This then triggers the creation of a \texttt{rdfs:comment} statement in the RDF graph, which is presented to the user whenever the resource is mouse hovered \circled{8}.

To derive a literal's datatype the following three checks are performed:
(1) If an input can be converted to an integer, the datatype of \texttt{xsd:int} is used.
(2) If the input is a floating-point number, the \texttt{xsd:float} type is assumed.
(3) Entering \textit{true} or \textit{false} triggers a \texttt{xsd:boolean}.
If all three cases do not apply the text is associated with a given language and stored as a \texttt{rdf:langString}.

As mentioned before, each resource is by default identified with a URI containing randomly generated UUID.
However, if an entered identifier can be assumed to be unique creating an additional URI is skipped.
For example, if a hyperlink like \texttt{https://iswc2017.semanticweb.org/} is entered, it is directly used as this resource's URI (without generating an additional one).

Correcting existing labels or literals is also possible by modifying their cell content.
This updates the corresponding \texttt{rdfs:label} or literal statement, respectively.
Setting the label's string to being empty (\texttt{""}) removes the link completely.
Note that this does not delete the resource itself (i.e., its corresponding triples) in order to be able to later use it in another spreadsheet, for example.

Further ideas for interesting future extensions will be addressed in Section \ref{sec:concl} after our evaluation and related work in the next sections.

\section{Evaluation}
\label{sec:eval}

We conducted a user study to show that our application is 1) easy to use and enables users to 2) create meaningful triple statements (even if they are rather inexperienced in RDF).

\paragraph{\textbf{Evaluation Setup}}
To estimate the ease of use of our application we measured the time to model a given scenario in RDF.
We learned from our industry projects that customers familiar with RDF often (ab)use Protégé\footnote{\url{http://protege.stanford.edu}} for creating triple statements.
Additionally, we also observed cases in which RDF is directly entered using a formal syntax like Turtle \cite{turtle}.
Thus, we compare our approach with these two in the following.

\begin{figure}
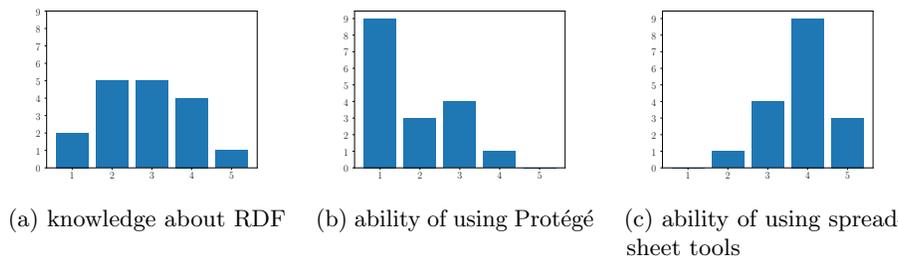

	\centering
	\begin{subfigure}[t]{0.3\linewidth}
		\resizebox{\linewidth}{!}{
				\input{img/eval-histogram-0.pgf}
		}
		\caption{knowledge about RDF}
		\label{fig:hist1}
	\end{subfigure}
   \quad
	\begin{subfigure}[t]{0.3\linewidth}
		\resizebox{\linewidth}{!}{
			\input{img/eval-histogram-1.pgf}
		}
		\caption{ability of using Protégé}
		\label{fig:hist2}
	\end{subfigure}
   \quad
	\begin{subfigure}[t]{0.3\linewidth}
		\resizebox{\linewidth}{!}{
			\input{img/eval-histogram-2.pgf}
		}
		\caption{ability of using spreadsheet tools}
		\label{fig:hist3}
	\end{subfigure}
	\caption{The participants' estimation of their knowledge and abilities on a 5-point scale ranging from \textit{novice (1)} to \textit{expert (5)}.}
	\label{eval-histo}
\end{figure}

We had 17 participants, 13 of them male, 4 female with an average age of 31.53 (standard deviation: 9.64).
To assess their knowledge about RDF, the usage of Protégé and spreadsheet tools, they were asked to estimate their skills on a 5-point scale ranging from \textit{novice (1)} to \textit{expert (5)}.
The results are summarized in the three histograms of Figure \ref{eval-histo}.
Apart from some novices and experts, most of the participants have mediocre RDF skills (i.e. RDF skills are approximately normal distributed).
In contrast, we had a large group of Protégé novices, no experts and just one that considered himself a near-expert.
In the case of spreadsheet tools, the distribution is quite opposite, 12 of 17 persons consider themselves experts or near-experts.

Looking at the three histograms, we see that lots of users are familiar with RDF without being an expert in Protégé, but they are quite experienced in using spreadsheet tools.

Each participant received the following task:
They are asked to develop an ontology by modelling the given information below as concisely as possible:\\

\par
\noindent
\begin{minipage}{\linewidth}
\leftskip=1cm
\textit{``Max attends the conference ESWC 2017.
The ESWC 2017 is located in Portoroz.
The keywords Semantic Web and Knowledge are related to ESWC 2017.
Portoroz is a city and lies within the country Slovenia.''}\\
\end{minipage}
\par

By giving the same scenario to all participants we intended to ensure that they model the same aspects.
Nevertheless, the results still vary between each of them due to existing freedom in modelling.
The very same information should be modelled in RDF by each participant using three different tools:
\begin{enumerate}
\item a common text editor of their choice using the Turtle syntax (TS),
\item using the ontology editor Protégé (P) in version 5.2.0,
\item using our RDF spreadsheet editor (RSE).
\end{enumerate}
To reduce learning effects over all participants we shuffled the order in which they used the applications, i.e. all permutations of (TS, P, RSE) occurred repeatedly.
Participants could use a text editor of their choice, which typically was Notepad++, Atom or Sublime Text.
They were encouraged to use usual shortcuts like copy \& paste, auto completion, etc.
Since writing Turtle was not part of their daily activities, they were given a small example from a different context that showed the syntax.
Additionally, they had a short tutorial on Protégé in the beginning and could even ask questions about its usage during the task.
The basic functionality of our application was briefly explained using a screen shot.

\paragraph{\textbf{Ease of Use}}
To estimate the ease of use, we measured for each application and each participant the time to create the respective triples of our scenario.
We started the time tracking with the user's first input and stopped it with their statement of being done with the task.
Besides the processing time we also counted the number of created RDF statements (\textit{stmts}).
If the submitted Turtle syntax could not be parsed, we counted the triples manually.
Due to syntax errors this was the case for 13 of 17 participants.
The results are depicted in Table \ref{time-table}.
\newcommand{\blanks}{~~~~~~}
\begin{table}[h]
\centering
\scriptsize
\begin{tabular}{|c||r|r||r|r||r|r|}
	\hline
	\multirow{2}{*}{Participant} & \multicolumn{2}{c||}{TS} & \multicolumn{2}{c||}{P} & \multicolumn{2}{c|}{RSE} \\
	&\blanks{} time &\blanks{}stmts &\blanks{}time &\blanks{}stmts &\blanks{}time &\blanks{}stmts \\
	\hline
	\hline
	01 &   719 & 40    & 534 & 27     & 370 & 60 \\
	02 &   527 & 42    & 1153 & 49     & 183 & 49 \\
	03 &   532 & 17    & 1458 & 36     & 693 & 53 \\
	04 &   327 & 20    & 602 & 30     & 178 & 64 \\
	05 &   589 & 28    & 641 & 35     & 193 & 49 \\
	06 &   377 & 42    & 486 & 33     & 187 & 49 \\
	07 &   1241 & 39    & 1066 & 35     & 198 & 52 \\
	08 &   567 & 14    & 419 & 22     & 228 & 41 \\
	09 &   568 & 22    & 1014 & 30     & 272 & 53 \\
	10 &   525 & 27    & 902 & 30     & 224 & 49 \\
	11 &   807 & 27    & 566 & 35     & 375 & 49 \\
	12 &   383 & 39    & 221 & 26     & 133 & 46 \\
	13 &   1423 & 40    & 625 & 46     & 368 & 56 \\
	14 &   734 & 21    & 605 & 32     & 321 & 69 \\
	15 &   858 & 38    & 531 & 30     & 202 & 46 \\
	16 &   384 & 19    & 378 & 29     & 124 & 46 \\
	17 &   589 & 35    & 335 & 27     & 209 & 49 \\
	\hline
	\hline
	\multicolumn{1}{|l||}{statements} & \multicolumn{2}{r||}{$\diameter 3.102$} & \multicolumn{2}{r||}{$\diameter 3.381$} & \multicolumn{2}{r|}{$\diameter 13.817$} \\
	\multicolumn{1}{|l||}{per minute} & \multicolumn{2}{r||}{$\pm 1.517$} & \multicolumn{2}{r||}{$\pm 1.369$} & \multicolumn{2}{r|}{$\pm 4.800$} \\
	\hline
\end{tabular}
\caption{Results of the user study: the time in seconds to finish the evaluation task (time), the number of created RDF statements (stmts) in the respective time interval and the average number (and standard deviation) of statements per minute for each of the three tools: writing turtle syntax (TS), using Protégé (P) and our RDF spreadsheet editor (RSE).}
\label{time-table}
\end{table}
To compare the results for the three different tasks we calculated the average number of created statements per minute (given at the bottom of Table \ref{time-table}).
In this comparison our approach achieves a score of $13.817$ statements per minute on average, which is four times the value of Protégé ($3.381$).
Although Protégé offers convenient possibilities to enter data, e.g. buttons and lists, several participants stated to be overwhelmed by the richness of the GUI.
Features to do bulk editing either seem to be non-existent or were not known to the participants -- a fact that may also be explained by the high number of Protégé novices in our study.
Writing Turtle in a text editor achieved similar results ($3.102$) as using Protégé.
In these editors the lack of a specialized graphical user interface is compensated by features like copy \& paste, text replacements, multi cursor, regular expressions, auto completion, etc.

\begin{figure}[h]
	\centering
	\scalebox{0.7}{
		\input{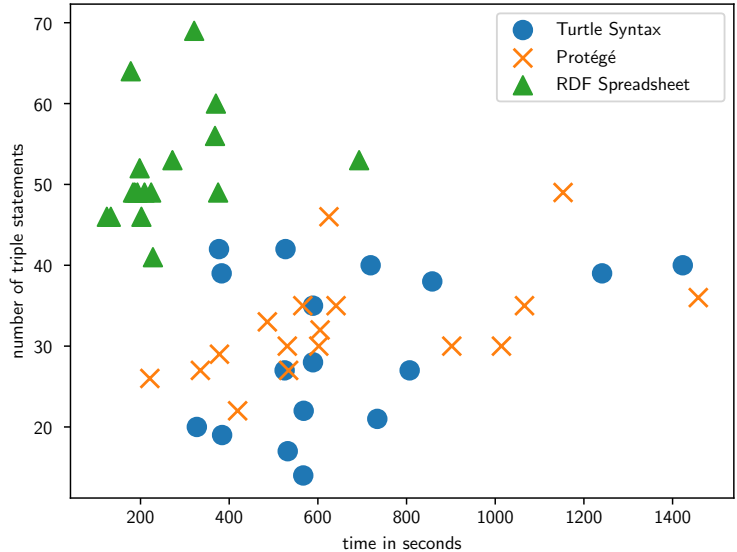}
	}
	\caption{Results of the user study: the time in seconds to finish the evaluation task (x-axis), the number of created RDF statements (y-axis) in the respective time interval  for each of the three tools: writing turtle syntax ($\circ$), Protégé ($\times$) and our approach ($\triangle$).}
	\label{eval-user}
\end{figure}

To give a visual impression of the results we transferred the data of Table \ref{time-table} to a scatterplot in Figure \ref{eval-user}.
The x-axis shows the time in seconds to complete a task, whereas the y-axis depicts the number of created triples in that time.
Each tool is represented with a different symbol: a text editor using turtle syntax ($\circ$), Protégé ($\times$) and our approach ($\triangle$).
Let us look at the results in more detail in the following, beginning with writing Turtle using a common text editor.
Although a few experienced users were able to create about 40 statements in a bit less than 400 seconds, most of the users were only able to create less than 40 in 400 to 800 seconds.
Two participants needed even longer (about 1200 and 1400 seconds).
Using Protégé resulted in similar results.
About 30 statements could be formulated in less than 600 seconds.
Again, some users needed more time (900 to 1400 seconds) for about the same number of statements.
The best result achieved was 46 statements in about 600 seconds.
In contrast, all users except for one were able to complete the tasks using our spreadsheet editor in less than 375 seconds (there was one participant that needed 693 seconds).
The resulting number of created statements was greater than 40 in all cases, whereas the best performing user was even able to create 69 statements in 321 seconds.

There is no clear performance difference between Protégé and Turtle, whereas using our application clearly led to the creation of more statements in less time.
We therefore consider our goal of creating an easy-to-use application to be fulfilled.

\begin{figure}[h]
	\centering
	\includegraphics[width=\textwidth]{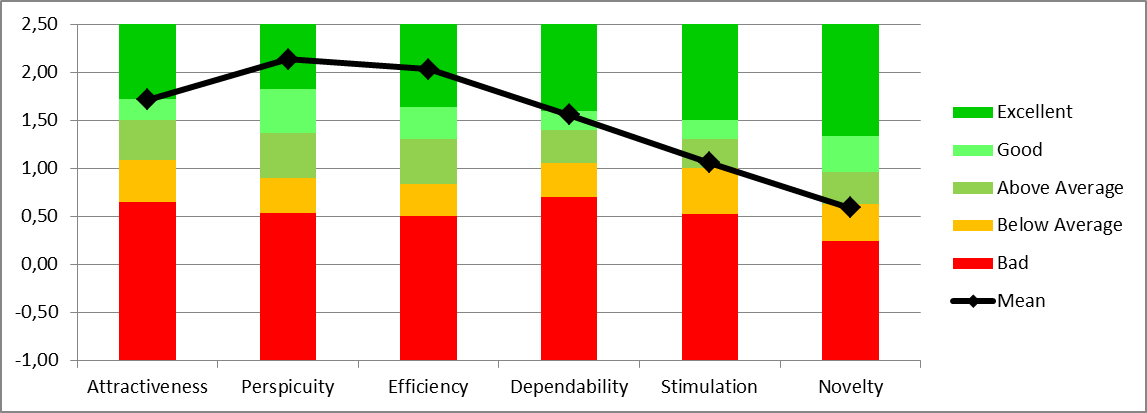}
	\caption{User experience questionnaire results}
	\label{ueq}
\end{figure}

To measure the user experience of our application we utilized the User Experience Questionaire (UEQ) proposed in \cite{ueqconstruction}.
Based on this questionnaire the following six factors are derived:
attractiveness, perspicuity, efficiency, dependability, stimulation, and novelty. 
Participants were asked to fill out the UEQ right after completing the task performed with our tool.
The results are depicted in Figure \ref{ueq}.
Our application receives excellent scores in attractiveness, efficiency and especially perspicuity.
The latter can be explained due to using the spreadsheet metaphor causing participants to operate on familiar terrain.
This familiarity aspect is also mirrored in the novelty scores, which are below average.
Thus, users did not perceive our application as being novel, which is exactly as intended.

Beside the ease of use, we also evaluated the creation of meaningful statements, which is addressed in the next section.

\paragraph{\textbf{Meaningful Triple Statements}}

Having an application that allows for easy RDF data creation would not be very useful if the resulting model is incorrect or of bad quality.

Concerning correctness, the results for each participant and each application were verified by the experimenter.
In all cases the given scenario has been modelled correctly beside some minor variations (due to personal design decisions).

To show that the ontologies resulting from using our tool are of similar quality than those created with Protégé, we calculated several quality metrics \cite{tartir2005ontoqa}.
In particular, we use the number of statements, classes, properties and instances as well as the relationship richness, attribute richness, class richness and average population.
Table \ref{metric-table} shows the resulting metrics.
Note that we omitted the Turtle results here, since most of them were not parsable.
For each metric we calculated the dependent t-test for paired samples with a significance level of 5\%.
Ontologies created with our application are significantly better with respect to the number of statements, instances, attribute richness, and average population.
For relationship richness Protégé slightly outperforms us, as the RDF spreadsheet editor creates more RDFS properties (like \texttt{rdfs:label}, \texttt{rdfs:domain} etc.) than Protégé (while the number of user-defined properties is about the same).
On the other hand, this causes the attribute richness to be notably higher, since each resource (especially classes) is associated with a label in our system.
The average population is roughly the same for both tools due to the approximately same number of modelled classes and instances in both cases.

In summary, we conclude that the results of our approach are at least comparable to those achieved by using Protégé, and significantly outperform it with respect to the number of statements, instances, attribute richness, and average population.

\begin{table}[h]
	\centering
	\scriptsize
	\begin{tabular}{|c|r|r|r|r|r|r|r|r|r|r|r|r|r|r|r|r|}
		\hline
		part. & \multicolumn{2}{c|}{stmt} & \multicolumn{2}{c|}{classes} & \multicolumn{2}{c|}{prop.} & \multicolumn{2}{c|}{inst.} & \multicolumn{2}{c|}{relationship} & \multicolumn{2}{c|}{attribute} & \multicolumn{2}{c|}{class} & \multicolumn{2}{c|}{average} \\
		 & \multicolumn{2}{c|}{} & \multicolumn{2}{c|}{} & \multicolumn{2}{c|}{} & \multicolumn{2}{c|}{} & \multicolumn{2}{c|}{richness} & \multicolumn{2}{c|}{richness} & \multicolumn{2}{c|}{richness} & \multicolumn{2}{c|}{population} \\
		&~~~~P& RSE  &~~~~P& RSE &~~~~P& RSE &~~~~P& RSE  &~~~~P& RSE &~~~~P& RSE&~~~~P& RSE            &~~~~P& RSE         \\
		\hline
		\hline
		01 & 27 & 60 & 5 & 5 & 4 & 5 & 6 & 6 & 0.185 & 0.150 & 0.000 & 1.000 & 1.000 & 1.000 & 1.200 & 1.200 \\
		02 & 49 & 49 & 5 & 5 & 4 & 4 & 6 & 6 & 0.102 & 0.102 & 0.800 & 1.000 & 1.000 & 1.000 & 1.200 & 1.200 \\
		03 & 36 & 53 & 6 & 6 & 4 & 4 & 4 & 6 & 0.167 & 0.094 & 0.000 & 1.000 & 0.667 & 0.667 & 0.667 & 1.000 \\
		04 & 30 & 64 & 3 & 3 & 5 & 8 & 4 & 7 & 0.200 & 0.125 & 0.000 & 1.000 & 1.000 & 1.000 & 1.333 & 2.333 \\
		05 & 35 & 49 & 5 & 5 & 4 & 4 & 6 & 6 & 0.143 & 0.102 & 0.000 & 1.000 & 1.000 & 1.000 & 1.200 & 1.200 \\
		06 & 33 & 49 & 5 & 5 & 4 & 4 & 6 & 6 & 0.152 & 0.102 & 0.000 & 1.000 & 1.000 & 1.000 & 1.200 & 1.200 \\
		07 & 35 & 52 & 5 & 5 & 4 & 5 & 6 & 6 & 0.143 & 0.115 & 0.000 & 1.000 & 1.000 & 0.800 & 1.200 & 1.200 \\
		08 & 22 & 41 & 4 & 4 & 3 & 3 & 4 & 6 & 0.227 & 0.122 & 0.000 & 1.000 & 1.000 & 1.000 & 1.000 & 1.500 \\
		09 & 30 & 53 & 5 & 5 & 5 & 4 & 6 & 7 & 0.200 & 0.113 & 0.000 & 1.000 & 1.000 & 1.000 & 1.200 & 1.400 \\
		10 & 30 & 49 & 6 & 5 & 5 & 4 & 6 & 6 & 0.133 & 0.102 & 0.000 & 1.000 & 0.833 & 1.000 & 1.000 & 1.200 \\
		11 & 35 & 49 & 5 & 5 & 4 & 4 & 6 & 6 & 0.143 & 0.102 & 0.000 & 1.000 & 1.000 & 1.000 & 1.200 & 1.200 \\
		12 & 26 & 46 & 5 & 5 & 3 & 3 & 6 & 6 & 0.192 & 0.109 & 0.000 & 1.000 & 1.000 & 1.000 & 1.200 & 1.200 \\
		13 & 46 & 56 & 6 & 6 & 5 & 4 & 7 & 8 & 0.152 & 0.089 & 0.000 & 1.000 & 1.000 & 1.000 & 1.167 & 1.333 \\
		14 & 32 & 69 & 4 & 4 & 7 & 8 & 4 & 6 & 0.219 & 0.145 & 0.000 & 1.500 & 1.000 & 1.000 & 1.000 & 1.500 \\
		15 & 30 & 46 & 5 & 4 & 5 & 5 & 4 & 4 & 0.167 & 0.130 & 0.000 & 1.000 & 0.800 & 1.000 & 0.800 & 1.000 \\
		16 & 29 & 46 & 6 & 5 & 3 & 3 & 6 & 6 & 0.172 & 0.109 & 0.000 & 1.000 & 0.833 & 1.000 & 1.000 & 1.200 \\
		17 & 27 & 49 & 5 & 5 & 4 & 4 & 6 & 6 & 0.185 & 0.102 & 0.000 & 1.000 & 1.000 & 1.000 & 1.200 & 1.200 \\
		\hline\hline
		avg. & 32.5 & 51.8 & 5.0 & 4.8 & 4.3 & 4.5 & 5.5 & 6.1 & 0.170 & 0.113 & 0.047 & 1.029 & 0.949 & 0.969 & 1.104 & 1.298\\
		s.d. & 6.8 & 7.1 & 0.8 & 0.7 & 1.0 & 1.5 & 1.0 & 0.8 & 0.033 & 0.017 & 0.194 & 0.121 & 0.101 & 0.092 & 0.171 & 0.300\\
		\hline
	\end{tabular}
	\caption{Ontology metrics \cite{tartir2005ontoqa} to compare the results of our approach (RSE) with ontologies generated by Protégé (P) for each participant (part.); used abbreviations: stmt: number of created RDF statements, prop.: properties, inst.: instances., avg.: average, s.d.: standard deviation}
	\label{metric-table}
\end{table}

\section{Related Work}
\label{sec:relwork}

While our approach combines the common spreadsheet metaphor with interactive and collaborative aspects, other related approaches exist.
For most of them a knowledge expert has to provide a mapping of how to transfer the spreadsheet data to RDF, whereas some approaches also support semi- or fully automated conversions without this necessity (see Section \ref{sec:mapping}).
Apart from taking filled spreadsheets as an input and converting them, other approaches focus on supporting the user in entering and working with RDF data (Section \ref{sec:rdfhelp}).
Since our application allows for entering data that is then transferred to RDF, we investigate the area of related work from two directions: existing RDF/ontology editors that may use the spreadsheet metaphor (Section \ref{sec:ontoedit}) as well as spreadsheet editors that may operate on RDF graphs (Section \ref{sec:ssedit}).

\subsection{Mapping \& Conversion}
\label{sec:mapping}

RDF123 \cite{han2008rdf123} uses an expressive mapping that has to be formulated in RDF.
It transfers rows into row graphs, which are then merged to get a full representation of the spreadsheet in RDF.
More complex mappings can be expressed using conditions, arithmetic and string manipulation.
The formulation of mappings induces additional effort and requires deeper knowledge about concepts of RDF.
Although the authors provide a graphical tool to support this process, basic RDF knowledge is needed nonetheless.

Sheet2RDF \cite{Fiorelli2015} follows a similar approach but focuses more on using heuristics to derive transformation rules automatically, which can then be refined by the user.

Spread2RDF\footnote{\url{https://github.com/marcelotto/spread2rdf}} also converts spreadsheets to RDF using a mapping language which is defined in Ruby.

R2RML (RDB to RDF Mapping Language) \cite{r2rml} and its extension RML (RDF Mapping Language) \cite{dimou2014rml} are very general mapping languages that transfer relational databases as well as structured data to RDF.

D2RQ \cite{bizer2004d2rq} also provides a mapping language but focuses on making relational databases queryable by SPARQL.

$M^2$ \cite{o2010m2} is a mapping language which is specialized for the transformation to OWL.

Sharma et al. \cite{sharmaautomatically} present an approach that automatically processes several spreadsheets in order to create an ontology (classes and properties) with instance data.
In this process the ranges of data type properties are extracted automatically, object type properties are found and classes are generated.

Any23 (Anything To Triples)\footnote{\url{https://any23.apache.org/}} is a library that is able to convert structured data, especially CSV, to RDF.
The algorithm converts every cell to a triple, whereas each row represents a resource and columns serve as properties.

csv2rdf4lod-automation\footnote{\url{https://github.com/timrdf/csv2rdf4lod-automation/wiki}} is an application that can also be used to convert CSV to RDF.

TabLinker\footnote{\url{https://github.com/Data2Semantics/TabLinker}} requires a specifically annotated Excel file in order to convert it to RDF.
It uses the built-in style functionality of Excel to save these annotations in the spreadsheet.

In summary, the previously presented approaches allow knowledge experts to create RDF data from various sources.
Typically a mapping language is used.
In contrast to our approach, the transformation steps are designed as a batch process and are not done simultaneously.
Additionally, the mappings needed for transformation are hard to define for inexperienced users.
Our approach uses a given and fixed mapping and does a simultaneous transformation.
None of the existing tools supports incremental updating.

\subsection{Support for Entering \& Processing RDF}
\label{sec:rdfhelp}

The following approaches focus on entering, processing and working with RDF data, especially in a user-friendly way.
Inexperienced users are therefore also able to create semantic data.

Pohl \cite{Pohl2014} published \textit{rdfedit}, a web-based tool to create RDF data that could also be used by Semantic Web laymen.
Using a subject-predicate-object table, users were able to enter labels which were mapped to URIs using the now discontinued Semantic Web search engine Sindice.
Auto completion of already loaded concepts was also possible.

RDForms\footnote{\url{http://rdforms.org/}} allows for creating RDF triples using forms which are predefined by templates.
Whenever the user enters something in the form, it is simultaneously transferred to RDF triples.
RDFEdit\footnote{\url{https://joinup.ec.europa.eu/software/rdfedit/description}} is an editor reusing RDForms which is able to search, view and edit RDF data.
Aditya Kalyanpur et al. \cite{kalyanpur2002rdf} proposed an RDF editor equipped with the RDF Instance Creator (RIC) feature.
They follow the same form-based approach like RDForms.

Another approach, RightField \cite{wolstencroft2011rightfield}, as its name suggests, focuses on ensuring that only data that is correct with respect to a given ontology may be entered into a spreadsheet.

Tripcel\footnote{\url{http://www.ifs.univie.ac.at/schandl/2009/06/tripcel}} also uses the spreadsheet metaphor to read and process RDF data.
Therefore, the author created an expression language which allows for defining RDF terms in a sheet.
Additionally, users may call functions on them just like they are used to from typical spreadsheet tools (e.g., sum function).

The previously mentioned approaches are more user friendly than typical ontology editors, which we will discuss in the next section.
Nevertheless, unlike our tool they still require considerable RDF knowledge in order to be used or configured.
None of them uses the spreadsheet metaphor in a way that enables a fast data entry as we do.

\subsection{Ontology Editors}
\label{sec:ontoedit}
Ontology editors are specialized on creating and modifying ontologies using a graphical user interface.
Since this requires certain experience in the knowledge modelling domain, they are intended to be used by experts than novices.
One of the most widely known editors is Protégé\footnote{\url{http://protege.stanford.edu}}, which is quite easily extensible due to its plug-in architecture.
Its GUI is adaptable to various modelling use cases.
Additionally, a web-based version called WebProtégé\footnote{\url{https://webprotege.stanford.edu/}} is available, that allows for collaboratively working on and sharing ontologies.
There are also commercial solutions that offer similar functionality like TopBraid Composer\footnote{\url{http://www.topquadrant.com/tools/ide-topbraid-composer-maestro-edition/}} by TopQuadrant or OntoStudio\footnote{\url{http://www.semafora-systems.com/en/products/ontostudio/}} by Semafora.

Swoop \cite{kalyanpur2006swoop} is an open-source and web-based ontology editor, which focuses on providing a web browser look and feel.
A text editor is used to modify the resources which can be ``browsed'' to using their URI.

The approaches presented in this section are well-suited for ontology creation, i.e. classes and properties (TBox).
They are rather intended for experts than for novices.
Regardless of whether a user is experienced or not, creating instances (ABox) is rather cumbersome with these editors.
But still, Protégé is often used as a standard data entry tool in industry.
Thus, our focus is on enabling users, especially novices, to easily create instance data.

\subsection{Spreadsheet Editors}
\label{sec:ssedit}

Probably the most widely known one is Microsoft Excel\footnote{\url{https://products.office.com/en-us/excel}} (commercial tool).
Open-source solutions are Apache OpenOffice Calc\footnote{\url{https://www.openoffice.org/product/calc.html}},
LibreOffice Calc\footnote{\url{https://www.libreoffice.org/discover/calc/}}
 and the web-based EtherCalc\footnote{\url{https://ethercalc.net/}}.
Other web-based solutions are WikiCalc\footnote{\url{http://www.softwaregarden.com/products/wikicalc/}}, SocialCalc\footnote{\url{https://www.socialtext.net/open/socialcalc}}, the Drupal\footnote{\url{http://www.drupal.com/}} module Sheetnode\footnote{\url{https://www.drupal.org/project/sheetnode}} and the commercial tool Zoho Sheet\footnote{\url{https://www.zoho.com/docs/sheet.html}}.

However, unlike our approach, none of these editors uses an RDF graph as basic data model and can easily export RDF without first defining a use-case specific mapping.

\section{Conclusion and Outlook}
\label{sec:concl}
In this paper we presented a collaborative web-based tool that uses the spreadsheet metaphor to enable all users, especially those inexperienced in Semantic Web concepts, to create RDF data.
Each entry into a cell of the spreadsheet is simultaneously transferred into triple statements.
This leads to the creation of four times as many statements in the same time compared to alternative approaches.
At the same time, the quality of the statements is similar to that achieved by using Protégé, and significantly outperforming it in four metrics.
An online demo of our application is available\footnote{\url{http://www.dfki.uni-kl.de/~mschroeder/rdf-spreadsheet-editor/}}.

In future, our investigations will focus on users creating and working on a knowledge base together, as well as creating and using a shared vocabulary.
To better support the entering process, we intend to directly visualize consequences of the current modelling step (e.g., introducing a new class or property).
This could be done using semantic graph visualization or showing previews of knowledge service results (e.g., faceted search).
Additionally, we intend to support generating spreadsheets pre-filled with imported structured data (especially existing knowledge graphs) that can then be modified or extended by the user.

\paragraph{\textbf{Acknowledgements}}
Parts of this work have been funded
by the German Federal Ministry of Economic Affairs and Energy
in the project PRO-OPT (01MD15004D)
and by the Deutsche Forschungsgemeinschaft (DFG, German Research Foundation)
in the project Managed Forgetting (DE 420/19-1).

\bibliographystyle{splncs03}
\bibliography{paper-stripped}

\end{document}